\documentclass[twocolumn]{aastex62}
\usepackage[version=4]{mhchem}
\usepackage{soul}

\received{ 2018 February 22}
\revised{2018 May 8}
\accepted{2018 May 9}
\submitjournal{AJ}
\shorttitle{UV Observations of CME Impact on 67P}
\shortauthors{Noonan et al.}
\begin{document}

\title{Ultraviolet observations of Coronal Mass Ejection impact on comet 67P/Churyumov-Gerasimenko by Rosetta Alice}

\author[0000-0003-2152-6987]{John W. Noonan}
\affiliation{Department of Space Studies,Southwest Research Institute, Suite 300, 1050 Walnut Street,Boulder, Colorado 80302,USA}
\affiliation{Lunar and Planetary Laboratory,University of Arizona, 1629 E University Blvd, Tucson, Arizona 85721-0092, USA }

\author{S. Alan Stern}
\affiliation{Department of Space Studies, Southwest Research Institute, Suite 300, 1050 Walnut Street, Boulder, Colorado 80302,USA}

\author[0000-0002-9318-259X]{Paul D. Feldman}
\affiliation{Department of Physics and Astronomy, The Johns Hopkins University, 3400 N. Charles Street, Baltimore, Maryland 21218, USA }

\author[0000-0001-6910-2724]{Thomas Broiles}
\affiliation{Space Science Institute. 4750 Walnut St, Suite 205, Boulder, Colorado 80301, USA}
 
 \author[0000-0003-2201-7615]{Cyril Simon Wedlund}
 \affiliation{Department of Physics, University of Oslo. Box 1048 Blindern, 0316 Oslo, Norway}
 
 \author[0000-0002-1261-7580]{Niklas J.T. Edberg}
 \affiliation{Swedish Institute of Space Physics, Lagerhyddsvagen 1, 75121, Uppsala, Sweden} 
 
 \author{Eric Schindhelm}
 \affiliation{Ball Aerospace and Technology Corp, 1600 Commerce St., Boulder, Colorado 80301, USA}
 
\author[0000-0002-3672-0603]{Joel Wm. Parker}
\affiliation{Department of Space Studies, Southwest Research Institute, Suite 300, 1050 Walnut Street, Boulder, Colorado 80302,USA}

\author[0000-0003-0797-5313]{Brian A. Keeney}
\affiliation{Department of Space Studies, Southwest Research Institute, Suite 300, 1050 Walnut Street, Boulder, Colorado 80302,USA}

\author[0000-0002-8227-9564]{Ronald J. Vervack Jr}
\affiliation{Johns Hopkins University Applied Physics Laboratory, 11100 Johns Hopkins Road, Laurel, Maryland 20723-6099,USA}

\author[0000-0002-5358-392X]{Andrew J. Steffl}
\affiliation{Department of Space Studies, Southwest Research Institute, Suite 300, 1050 Walnut Street, Boulder, Colorado 80302,USA}

\author[0000-0003-2781-6897]{Matthew M. Knight}
\affiliation{Astronomy Department, University of Maryland, College Park, Maryland 20742, USA}

\author[0000-0003-0951-7762]{Harold A. Weaver}
\affiliation{Johns Hopkins University Applied Physics Laboratory, 11100 Johns Hopkins Road, Laurel, Maryland 20723-6099,USA}

\author[0000-0002-4230-6759]{Lori M. Feaga}
\affil{Astronomy Department, University of Maryland, College Park, Maryland 20742, USA}

\author{Michael A'Hearn}
\altaffiliation{\textit{Deceased}}
\affiliation{Astronomy Department, University of Maryland, College Park, Maryland 20742, USA}

\author{Jean-Loup Bertaux}
\affiliation{LATMOS, CNRS/UVSQ/IPSL, 11 Boulevard d'Alembert, 78280 Guyancourt, France}

\correspondingauthor{John Noonan}
\email{noonan@boulder.swri.edu}

\begin{abstract}

The Alice ultraviolet spectrograph on the European Space Agency \textit{Rosetta} spacecraft observed comet 67P/Churyumov-Gerasimenko in its orbit around the Sun for just over two years. Alice observations taken in 2015 October, two months after perihelion, show large increases in the comet's Lyman-$\beta$, \ion{O}{1}~1304, \ion{O}{1}~1356, and \ion{C}{1}~1657~\AA\ atomic emission that initially appeared to indicate gaseous outbursts. However, the \textit{Rosetta} Plasma Consortium (RPC) instruments showed a coronal mass ejection (CME) impact at the comet coincident with the emission increases, suggesting that the CME impact may have been the cause of the increased emission. The presence of the semi-forbidden \ion{O}{1}~1356~\AA\ emission multiplet is indicative of a substantial increase in dissociative electron impact emission from the coma, suggesting a change in the electron population during the CME impact. The increase in dissociative electron impact could be a result of the interaction between the CME and the coma of 67P or an outburst coincident with the arrival of the CME. The observed dissociative electron impact emission during this period is used to characterize the \ce{O2} content of the coma at two peaks during the CME arrival. The mechanism that could cause the relationship between the CME and UV emission brightness is not well constrained, but we present several hypotheses to explain the correlation. 
\end{abstract}

\keywords{comets---individual(67P/C-G), Sun---coronal mass ejections, ultraviolet---planetary systems}
%%%%%%%%%%%%%%%%%%%%%%%%%%%%%%%%%%%%%%%%%%%%%%%%%%

%%%%%%%%%%%%%%%%% BODY OF PAPER %%%%%%%%%%%%%%%%%%

\section{Introduction}\label{introduction}

The European Space Agency (ESA) \textit{Rosetta} spacecraft was launched in 2004 to perform an orbital study of the comet 67P/Churyumov-Gerasimenko, the first mission of its kind. Following rendezvous with the comet on 2014 August 6, the \textit{Rosetta} spacecraft was able to observe the surface and activity of the comet from close distances. The Alice ultraviolet spectrograph on board the spacecraft measured the atomic and molecular far ultraviolet (FUV) emissions. These observations help to characterize the atomic and molecular composition, reflectance properties of the comet's surface and the composition and time variation of the comet's coma \citep{stern2007alice}. 

Previous papers analyzing Alice data have explored the near nucleus coma (d$_{comet}$ $\leq$ 100~km) environment, the dominant emission from electron impact dissociation of water, and the spectral signature of outbursts from the nucleus \citep{stern2015first,feldman2015measurements,feldman2016nature,chaufray2017rosetta,keeney2017h2o,feldman2018fuv}. These studies have shown that the contribution of \textcolor{black}{dissociative electron impact excitation to coma emission} is significant and observable, as well as that molecular oxygen (\ce{O2}) appears to be abundant, even more so than pre-perihelion in-situ mass spectrometer data have shown \citep{bieler2015abundant,fougere2016three}.

The interaction between solar system objects and powerful solar events like coronal mass ejections (CMEs) has long been a subject of interest for space physicists and planetary scientists alike. Emission spikes in conjunction with the arrival of solar events have been observed on other solar system objects as well, though none as small as a comet. For example, observations of Venus' atmosphere during solar events showed a substantial increase to the \ion{O}{1} 5577~\AA\ emission line following interactions with CMEs, co-rotating interaction regions (CIRs), or the solar wind \citep{gray2014effect}. Substantial data have been gathered on both Earth's and Mars' ionospheric reactions to CME impacts indicating that a CME arrival is accompanied by a compression of the planetary magnetosphere, precipitation of energetic particles into the atmosphere, and an increase in electron density, as well as aurora and nightglow emission \citep{GRL:GRL26046}. Additionally, modeling of the Martian atmosphere has shown that during a solar energetic particle event the electron density could reach as high as 10$^{4}$ cm$^{-3}$ within 100~km of the surface \citep{sheel2012numerical}. A combined CME/CIR impact occurring 2014 October 22 on 67P was observed and described in \citet{edberg2016solar} and \citet{witasse2017interplanetary} and witnessed by Alice. The resulting emissions are described in \citet{feldman2015measurements} but were only recognized as the result of the CME impact following that paper's publication. Visible observations of the 2007 CME impact on comet 2P/Encke by STEREO were from too great a distance to directly observe the behavior of the inner coma in response to the increased flux of energetic particles \citep{vourlidas2007first}. 
 
Here we describe a substantial increase in atomic UV emission lines coincident with the arrival of a coronal mass ejection (CME) at 67P. A brief overview of the Alice instrument is given in Section 2. Section 3 discusses observations of the observed emission spike that was seen on 2015 October 5-6, the same date and time that \cite{doi:10.1093/mnras/stw2112} reported that a CME impacted the coma of 67P. Section 4 reviews the Alice spectra and Ion and Electron Spectrometer (IES) data gathered during this CME, and compares Alice and Rosetta Plasma Consortium (RPC) results to establish that there is a relationship between the impact of the CME and the UV emissions observed by Alice. In Section 5 we discuss the possible sources of electrons that could contribute to the observed emission and discuss the \ce{O2}/\ce{H2O} ratio calculated during the emission spikes. Section 6 provides a summary. 

\section{The Alice Spectrograph}\label{instrument}
The Alice Spectrograph was designed to characterize the surface, coma, and nucleus/coma coupling of comet 67P. 
It is a low-power, lightweight, imaging far-ultraviolet (FUV) spectrograph designed to gather spatially resolved spectra from 700--2050~\AA\ with a spectral resolution of 8--12~\AA\ for sources that extend the length of the slit \citep{stern2007alice}.
The rectangular slit is $5.5^{\circ}$ long and has a shape reminiscent of a ``dog bone", wider on the bottom and top than the middle.  The slit is $0.05^{\circ}$ (100 $\mu$m) wide in the middle $2.0^{\circ}$ section of the slit, and $0.10^{\circ}$ (210 $\mu$m) wide in the top and bottom sections for a spectral resolution of 8~\AA\ and 12~\AA\, respectively. The microchannel plate (MCP) detector active area is 35$\times$20 mm with a pixel size 34$\times$620 $\mu$m for the 1032 spectral columns and 32 spatial rows. Rows 6-24 are illuminated by the slit, the other rows only see dark counts. \textcolor{black}{Detector rows 12 and 18 are transition rows with intermediate solid angles between the narrow and wide sections}. Each detector row subtends $0.30^{\circ}$ on the sky. The detector has two solar blind photocathodes (CsI and KBr) and a two-dimensional double delay-line readout enabling spectral and spatial information to be logged for every detected photon. The system uses an off-axis telescope feeding into a 0.15-m normal incidence Rowland circle spectrograph with a concave holographic reflection grating, all in an open environment. At the comet, the system experienced an unexpected time-variable feature blue-ward of Lyman $\beta$ between columns 700 and 900 on the detector, most likely due to cometary dust and ions impacting the detector \citep{noonan2016investigation}. The feature did not affect the analyses presented in this paper. 

\begin{figure}
\centering
\includegraphics[clip,trim=3cm 5cm 3cm 3cm,width=1.0\linewidth]{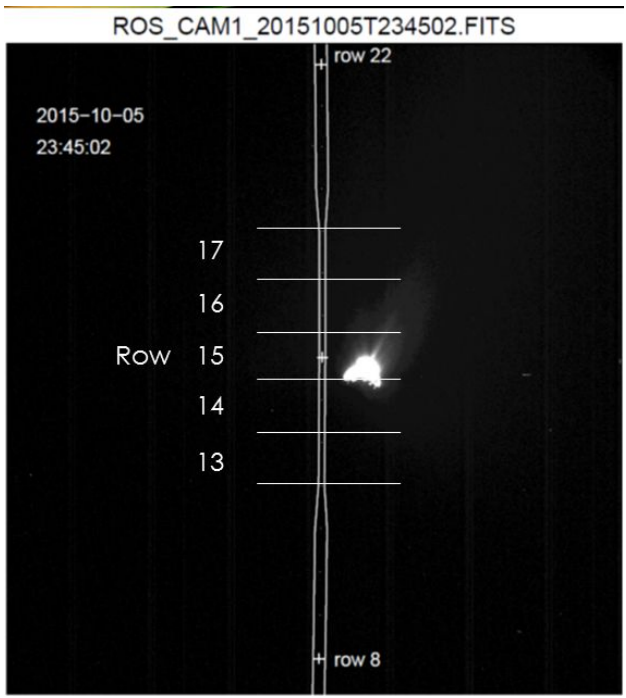} 
\caption{A \textit{Rosetta} NavCam image taken on 2015 October 5 at 23:45:02, just prior to the CME impact, with the Alice slit overlaid. The Sun is toward the top, illuminating portions of both the head and body of 67P. The flattest ``underside" portion of the body is facing \textit{Rosetta}. At this time the full length of the Alice slit subtends 76~km at the nucleus distance, approximately 4.2~km per pixel. (Image Credit: NAVCAM) }
\label{fig:1005_NavCam}
\end{figure}

\section{Observations and Analysis}\label{observations}
The emission spikes we discuss in this paper were observed during 23:30--03:30 UTC 2015 October 5--6 and were captured by the Alice instrument during observing schemes that were not designed for optimal characterization of such activity. The large cometocentric distance of \textit{Rosetta} means that the UV emission is sampled from an area much closer to the nucleus than the spacecraft. Due to the less than optimal pointing of Alice for this period, it is useful to review the observation design and pointing scenario. 

\subsection{Alice Observations}\label{raliceobs}
The Alice instrument has multiple observation modes, the most common being a five or ten-minute ``histogram" that uses the double delay-line detector to integrate a 2-D wavelength and spatial position image, where each pixel is a sum of the events detected at that spatial-spectral location \citep{stern2007alice}. This observing mode is optimal when Alice's slit is stationary relative to its target. Any scanning motion of the slit in the along-slit direction at a rate greater than one spatial pixel per exposure time will smear the image. The Alice data files contain SPICE-based pointing and geometry information at the start of the exposure, but no information about how the pointing changes during the exposure.

\begin{figure*}
\includegraphics[width = 1.0\linewidth]{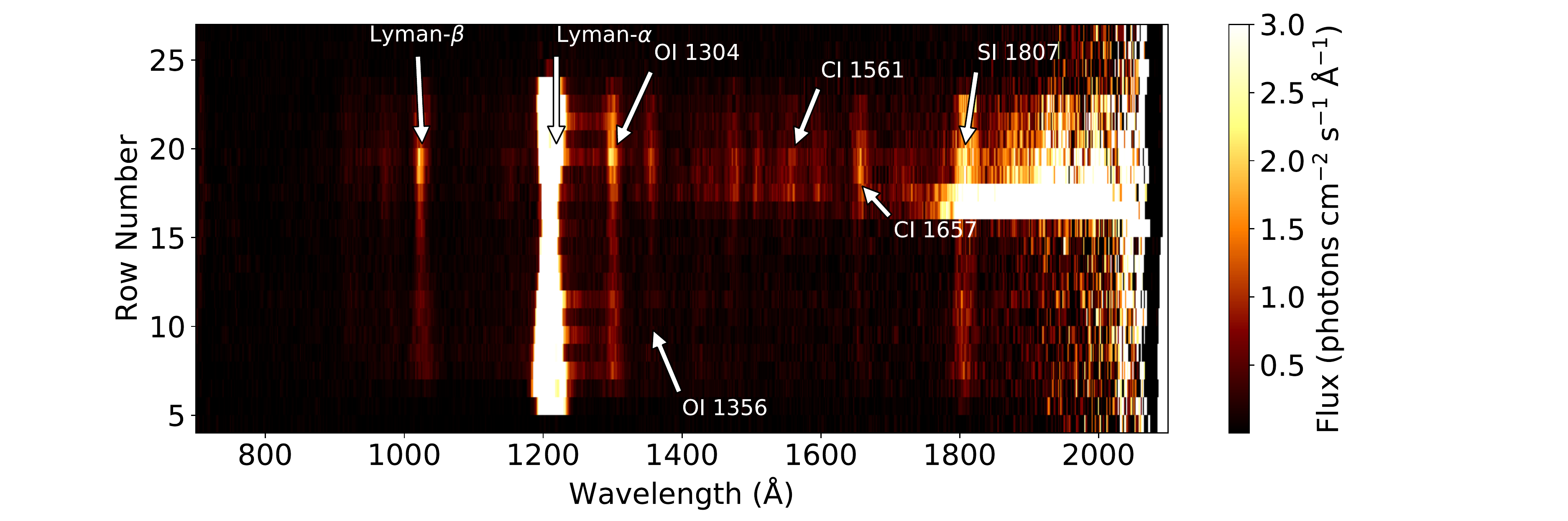}
\includegraphics[width = 1.0\linewidth]{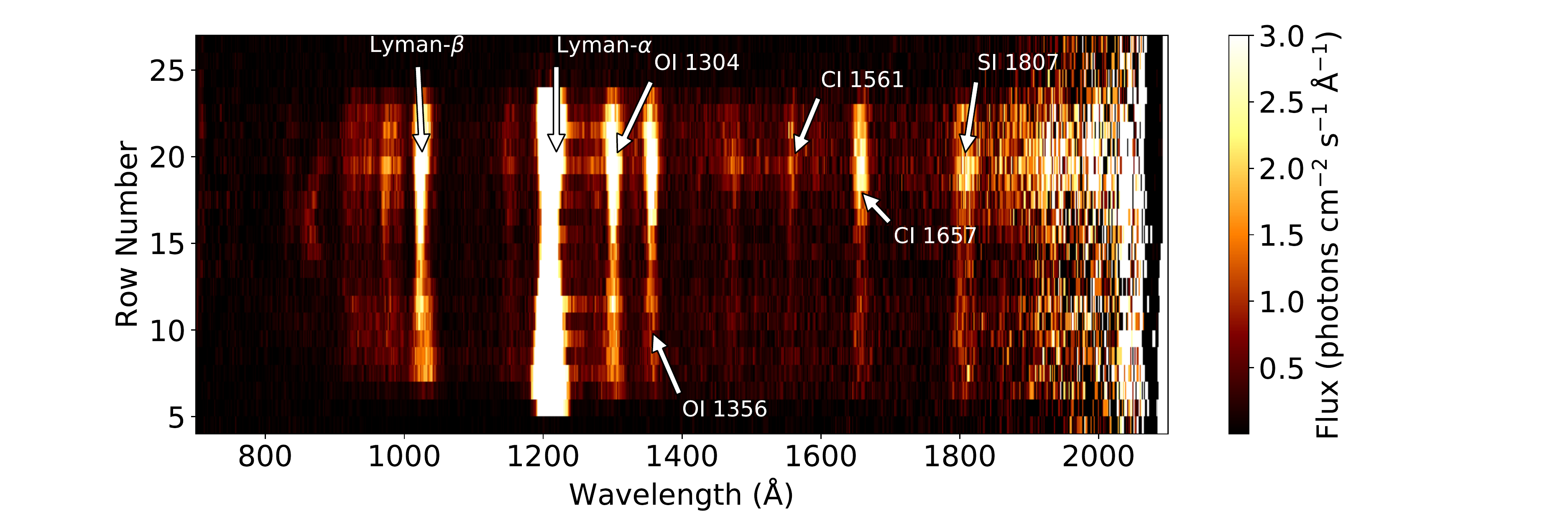}
%\plottwo{f2a}{f2b}
\caption{Top: Alice spectrum before CME impact taken at 00:38:59 UTC 5 October with similar, though not identical, pointing to the CME impact period. Reflected sunlight from the nucleus can be seen in the 1700--2100~\AA\ area of rows 17 and 18. Bottom: Alice spectrum during CME impact taken at 00:17:38 UTC 6 October. }
\label{fig:1005_spectra}
\end{figure*}

\begin{figure*}
\includegraphics[width = 1.0\linewidth]{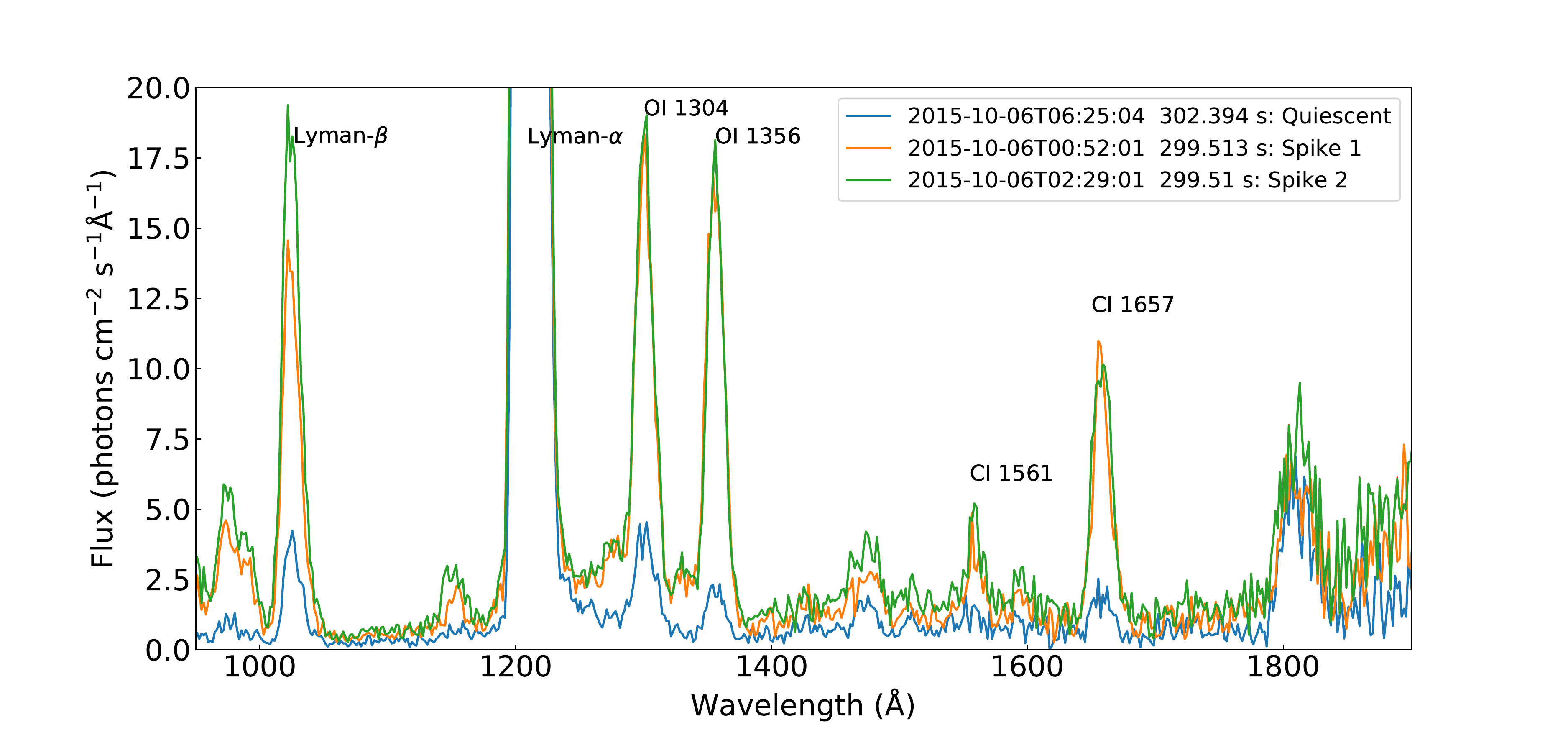}
\includegraphics[width = 1.0\linewidth]{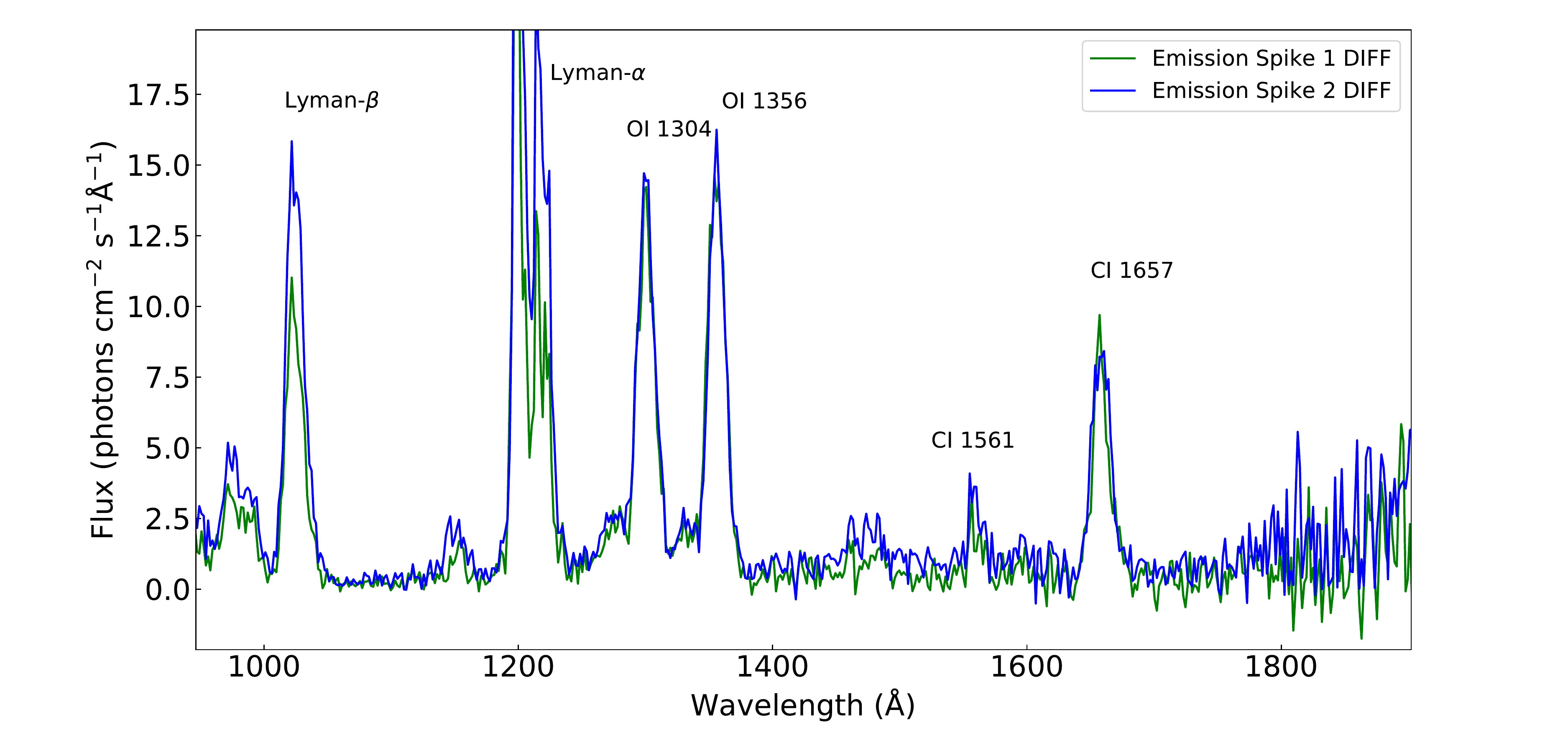}
\caption{ Top: Three spectra taken by the Alice instrument during similar pointing instances but with three distinct emission signatures. All spectra are made using rows 13-17, representing the rows closest to the nucleus. Integration time for each image is stated in the legend. \textcolor{black}{Statistical uncertainties are plotted but are smaller than the line thickness.} Bottom: The first and second emission spikes with the quiescent spectrum subtracted are plotted. Notice the increase in Ly-$\beta$ emission between the first and second emission spikes. } 
\label{fig:CME_Spectra}
\end{figure*}

Resonance fluorescence and dissociative electron impact on gases are expected sources of emission multiplets measured by Alice during this post-perihelion observation. Electron impact is believed to be the more significant source of emission for the period analyzed in this paper due to the dominance of the semi-forbidden \ion{O}{1}~1356~\AA\ multiplet (discussed further below). Experimentally determined electron impact emission efficiencies, or cross sections, as a function of energy for Ly-$\beta$, \ion{O}{1}~1304~\AA, \ion{O}{1}~1356~\AA, and \ion{C}{1}~1657~\AA\ emission multiplets are used to constrain the composition of the coma. This is done by comparing observed line ratios to the ratios of the 100 eV cross sections for emission features for qualitative gas compositions displayed in Table \ref{table:1}. There is some tolerance to the values given in Table \ref{table:1} due to the variation in the average electron energy at the comet, but in general the cross section ratios taken at 100 eV are best characterized in the literature. The ratios of the energy integrated cross sections provided in section \ref{electronimpact} are similar to the ratios of the cross sections at 100 eV. Table \ref{table:1} and other multiplet emissions are used to analyze the UV spectra gathered during the CME in Section \ref{results}. While not ideal, mixed gas electron impact UV emission studies have yet to be attempted. 

\subsection{Geometry and Spacecraft Pointing}\label{geometry}
Starting at 23:30 UTC 2015 October 5 and continuing to 06:00 October 6, the \textit{Rosetta} remote sensing instruments were performing a steady off-nadir angle scan. \textcolor{black}{The Rosetta spacecraft was in an approximately terminator orbit at 40$^{\circ}$ latitude on the nucleus.} The comet orientation for this period of time is captured in the NavCam image in Figure~\ref{fig:1005_NavCam}. At this time \textit{Rosetta} was at a heliocentric distance of 1.4 AU, having reached perihelion on 2015 August 13. The spacecraft was on its way back to the near-nucleus coma from a day-side excursion that took place in late 2015 September. During this excursion \textit{Rosetta} reached 1500~km from the nucleus in the Sun-ward direction. The CME impact occurred when \textit{Rosetta} was traveling towards the nucleus, from distances between 800 and 750 km \citep{doi:10.1093/mnras/stw2112}.

Observations from VIRTIS \citep{coradini2007virtis} post perihelion place \ce{H2O} column densities around \textcolor{black}{2-4 $\times 10^{20}$ m$^{-2}$ at this time \citep{bockelee2016evolution}}. Alice observations of the water column density in the months leading up to perihelion support the observations made by VIRTIS \citep{chaufray2017rosetta}. 

\begin{table}
\centering
\begin{tabular}{|| c c c c ||}
 \hline
 Gas & $\frac{OI~1304}{OI~1356}$ & $\frac{CI~1657}{HI~Ly-\beta}$ & $\frac{HI~Ly-\beta}{OI~1304}$ \\[.5ex]
 \hline
 H$_{2}$O & $\sim$3 & 0 & $\sim$3\\
 CO$_{2}$ & $\sim$2 & $\sim$1 & 0\\
 CO$_{2}$ and O$_{2}$ & $\sim$1 & $\sim$3 & 0\\
 O$_{2}$ & 0.5 & 0 & 0\\[1ex] 
 \hline
\end{tabular}
\caption{Electron impact emission line ratios for various gases and qualitative compositions relevant to cometary activity derived from \protect\cite{ajello1971dissociative}, \protect\cite{ajello1971emission}, \protect\cite{makarov2004kinetic}, \protect\cite{kanik2003electron}, and \protect\cite{mumma1972dissociative}}
\label{table:1}
\end{table}

The scanning motion of the Alice instrument was along the Sun/comet line, parallel to the direction of the slit. Over the course of the impact observations the scanning motion of the slit ranged from 0.00006 to 0.03 degrees per second, reducing the effectiveness of plotting the emission as a function of row/distance from nucleus. For the observations closest in time to the occurrence of the emission spikes discussed in this paper, the scanning rate averaged 0.001 degrees per second, or about one detector row per 300 second observation. These observations took place with an off nadir angle less than one degree from the nucleus, allowing a line of sight that captures emission from the near-nucleus environment, under 10 km from the nucleus.  

Just after 00:00 UTC on October 6, the brightness of all measured emission lines began to increase (Figure~\ref{fig:1005_spectra}). \ion{O}{1}~1304 and 1356~\AA\ has a relatively uniform brightness across the slit, while \ion{C}{1} 1657 \AA\ and weak \ion{C}{1} 1561 \AA\ are present in the upper rows as well. \textcolor{black}{ Note the appearance of weak sulfur and carbon multiplets at 1429 and $\sim$1470~\AA\ , respectively, and a stronger sulfur multiplet at 1807, 1820, and 1826~\AA}. The presence of sulfur and sulfur-bearing species in the coma has been reported by \citet{calmonte2016sulphur}. Several spatially summed spectra from detector rows 13-17 for this time are shown in Fig.~\ref{fig:CME_Spectra} to display the unique emission observed during this period. The first observation, taken on October 5 at 00:49 and plotted in blue, shows the coma two rotations ($\sim$24 hrs) prior to the CME impact from a similar off-nadir angle and Sun/comet orientation at a cometocentric distance of 860~km. The nucleus is captured as well in these early observations, producing the continuum at the red-ward side of the detector. The second observation, taken on October 6 at 00:52 and plotted in orange, shows a spectrum taken during the first spike of emission in Alice data. The third observation, taken on October 6 at 02:29 and plotted in green, is from the second emission spike. The second and third spectra are from the two spikes of the distinct increases in emission. The emission values of \ion{O}{1}~1304~\AA\ are nearly identical for the two spikes. In contrast, the second spike of the Lyman-$\beta$ emission line is significantly stronger than the first and both \ion{O}{1}~1356~\AA\ and \ion{C}{1}~1657~\AA\ are weaker. All observations have a similar pointing scheme and $\leq$ 1$^{\circ}$ off-nadir angle for the center of the slit. The oxygen, carbon, and hydrogen emissions for the first period have similar relative increases, but only the hydrogen emission increases further in the second emission spike. When the quiescent coma spectrum is subtracted from the emission spike spectra to produce a difference spectrum, this relative change is more pronounced (Fig.~\ref{fig:CME_Spectra}).

By integrating the emission multiplets for spectra taken between the 5th and 6th of October where the slit center, which coincides with detector row 15, is within 1~degree of the nucleus a plot of their brightness as a function of time can be used to look for the key moments and areas of emission. This is shown in Figure~\ref{fig:Smoothed_October}. Each observed multiplet experiences two emission spikes; the secondary spike for Ly-$\beta$ is stronger than the primary. This is in contrast to the other multiplets where the secondary emission peaks are weaker than the primary. Each multiplet experiences a relatively smooth decrease back to quiescent levels starting on October 6th at 03:15. It should be noted that the line of sight for the Alice instrument did not intersect the nucleus of 67P during the CME impact period, except for the set of observations made October 5 between 21:11 and 21:39 UTC and one observation on October 6 at 02:48 UTC. This means, as with all limb or coma observations, that the interplanetary medium Lyman-$\alpha$ and $\beta$ emissions are included in the observations at a constant background level that are subtracted in the the quiescent-subtracted spectra for compositional analysis. Compared to the observations taken 24 hours (i.e. approximately two comet rotations) earlier from a distance of 860~km, we see that line brightnesses increased 5--8 times for Ly-$\beta$ and \ion{O}{1}~1304~\AA, and approximately 15 times their quiescent value for \ion{O}{1}~1356~\AA\ (Figure~\ref{fig:Smoothed_October}).

\subsection{Complementary Observations}\label{complementaryobs}
To correlate Alice observations to the CME passage we use \textit{in situ} data gathered by the Rosetta Plasma Consortium instruments, specifically the Ion and Electron Spectrometer (RPC-IES) \citep{burch2007rpc}. During the CME impact RPC-IES collected data on the electron energy distribution at \textit{Rosetta} at regular intervals \citep{doi:10.1093/mnras/stw2112}, and these data that have been fit with kappa functions described in \cite{JGRA:JGRA52830}. RPC-IES measures electrons above 4.3 eV, which allows measurement of the lowest energy electrons responsible for dissociative electron impact emission. Threshold energies for dissociative electron impact are unique to each molecule, but the lowest threshold energies are $\sim$15 eV for relevant UV emission features. RPC-IES measurements can characterize all electrons that can contribute to the electron impact emission features with minimal effect from the spacecraft's potential, which is typically negative, and therefore repels a portion of the low energy electrons from IES below the threshold energies. Suprathermal electrons (10--200 eV) may have energies linearly shifted, but this effect is small \citep{clark2015suprathermal}. Observations from the other four RPC instruments are not discussed in detail but are mentioned in this paper. 

\subsection{Dissociative Electron Impact Analysis}\label{electronimpact}
Laboratory experiments have measured the cross sections for the \ion{O}{1}~1304~\AA, \ion{O}{1}~1356~\AA, \ion{C}{1} 1657~\AA, and Ly-$\beta$ transitions for electron impact on each of the expected major components of the coma: \ce{H2O}, \ce{CO2}, \ce{O2}, and \ce{CO} \citep{ajello1971dissociative,ajello1971emission,makarov2004kinetic,mumma1972dissociative,kanik2003electron,ajello1985study,ajello1971emission_co}. The four dominant molecules in the coma can dissociate into O fragments and be excited into OI, allowing for \ion{O}{1}~1304~\AA\ and \ion{O}{1}~1356~\AA\ from all four sources. The molecule- and transition-specific cross sections for Ly-$\beta$ and \ion{C}{1}~1657~\AA\ are used as indicators for \ce{H2O} and \ce{CO2}, respectively. Ly-$\alpha$ is not used for this analysis due to instrument gain sag in that portion of the detector, though relative changes are still apparent. 

Mathematically, the ratio of the \ce{O2} and \ce{H2O} column densities can be written as a function of the observed brightnesses of the \ion{O}{1}~1304~\AA, \ion{O}{1}~1356~\AA, and Ly-$\beta$ in the coma and their energy integrated cross sections:
\begin{equation}
\frac{N_{\ce{O2}}}{N_{\ce{H2O}}} = 0.068 \bigg( \frac{B_{OI 1304,\textnormal{Total}}}{B_{OI 1304,\ce{H2O}}} -\frac{B_{OI 1304,\ce{CO2}}+B_{{OI 1304,\ce{H2O}}}}{B_{OI 1304,\ce{H2O}}} \bigg)
\label{eq:1}
\end{equation}
\begin{equation}
\frac{N_{\ce{O2}}}{N_{\ce{H2O}}} = 0.104 \bigg( \frac{B_{OI 1356,\textnormal{Total}}}{B_{Ly-\beta,\ce{H2O}}} -\frac{B_{OI 1356,\ce{CO2}}+B_{{OI 1356,\ce{H2O}}}}{B_{Ly-\beta,\ce{H2O}}} \bigg)
\label{eq:2}
\end{equation}

\noindent where \textit{N} is the column density, \textit{B} is the brightness of the emission feature in Rayleighs, and the numerical constant represents the ratio of energy integrated g-factors for the \ion{O}{1}~1304~\AA\ and \ion{O}{1}~1356~\AA\ features between \ce{O2} and \ce{H2O}. Each individual integrated g-factor can be calculated using:
\begin{equation}
G_{\lambda,y} = \int_{E_{T}^{y}}^{300eV}\sigma_{\lambda}^{\ce{y}}(E)f_{pde}(E)dE,
\label{eq:3}
\end{equation}
\noindent where $\sigma$ is the analytically derived cross section efficiency of dissociative electron impact for molecule \textit{y} and emission feature $\lambda$ as a function of electron energy as described in \citet{shirai2001analytic} and \citet{kanik2003electron}. The lower limit of integration E$_{T}$ is defined as the threshold energy below which the emission feature will not appear. \textit{f$_{pde}(E)$} is the electron population distribution as measured at the spacecraft. 

This method loses effectiveness during periods with small amounts of electron impact activity, which are associated with a \ion{O}{1}~1304/1356 \AA\ ratio near 1. \ce{CO2} and \ce{CO} have \ion{O}{1}~1304~\AA\ and \ion{O}{1}~1356~\AA\ cross sections similar to \ce{O2}, but have dissociative cross sections for unique carbon emission features, producing a ``fingerprint" in spectra at the \ion{C}{2}~1335 \AA, \ion{C}{1}~1561 \AA, and \ion{C}{1}~1657~\AA\ multiplets; if seen, these would indicate that \ce{CO2} and/or \ce{CO} are present rather than \ce{O2}. The low \ce{CO}/\ce{CO2} ratio observed at 67P around perihelion \citep{mall2016high} and similarity of the carbon features to dissociative electron impact of \ce{CO2} makes \ce{CO} unlikely to contribute significantly to these carbon multiplets for the CME impact period. For this reason \ce{CO} is excluded from subtraction in Equations \ref{eq:1} and \ref{eq:2}.

\begin{figure*}
\includegraphics[clip,trim=0cm 3cm 0cm 3cm,width=1.0\linewidth]{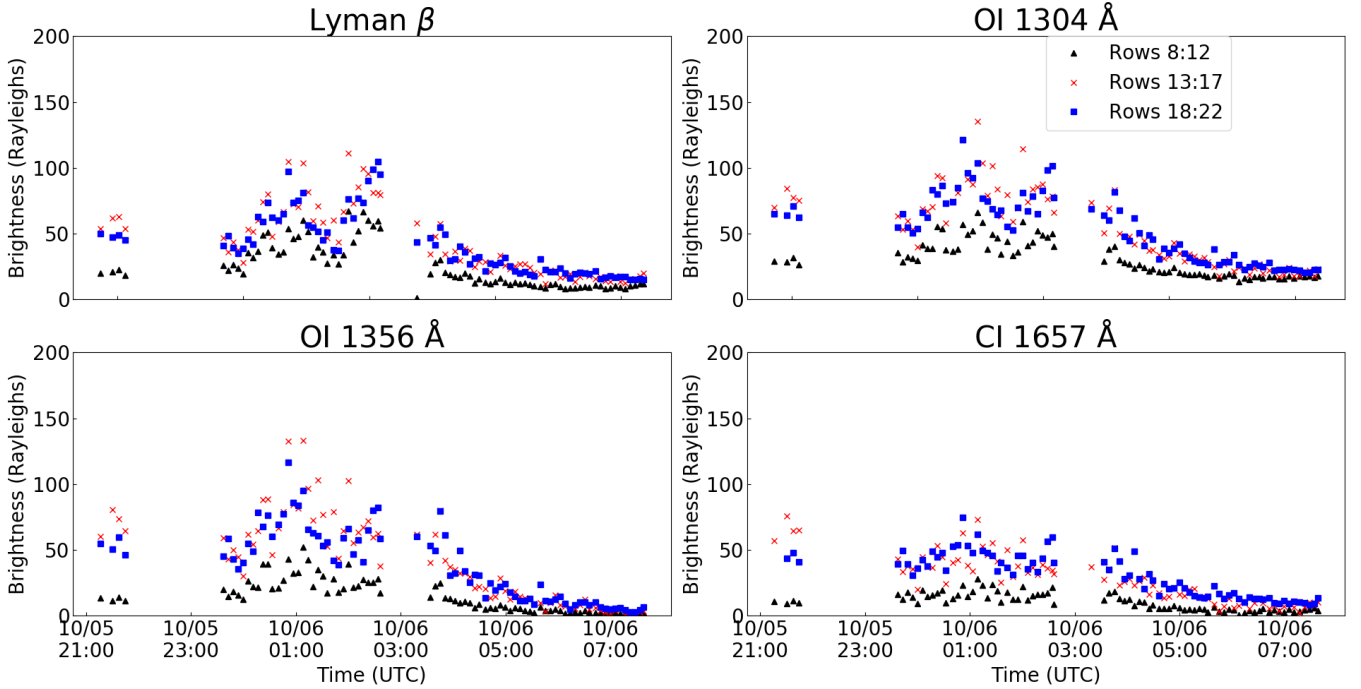}
\caption{Brightness vs. Time for October 5--6. All times displayed are in UTC. Gaps in data indicate periods where the Alice slit was more than 1 degree off of nadir or was not taking data. The nucleus is closest to rows 13--17 for this period, with rows 18--22 capturing sunward coma and rows 8--12 capturing anti-sunward coma. Rows closest to the nucleus see the strongest emission, followed by the sunward and anti-sunward coma. The solid angle differences for rows 12 and 18 are corrected for in the brightness calculation. The largest relative increase in emission occurs for \ion{O}{1}~1356~\AA. 1-$\sigma$ error bars are not plotted but are between 1--7 Rayleighs.}
\label{fig:Smoothed_October}
\end{figure*}

\begin{figure}
\includegraphics[width=\linewidth]{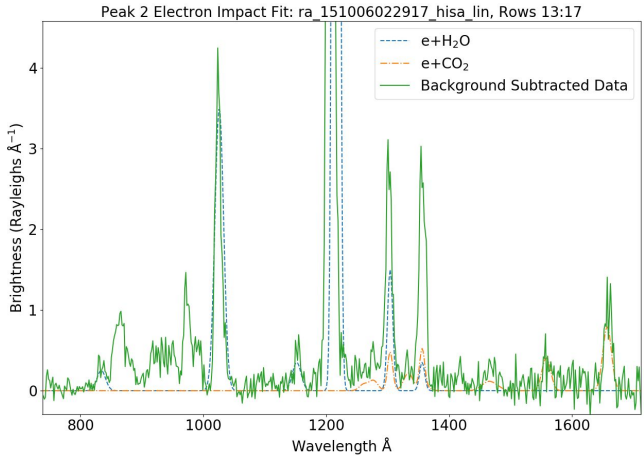}
\caption{The quiescent and background-subtracted spectrum from the second spike with dissociative electron impact of \ce{H2O} and \ce{CO2} (e+\ce{H2O} and e+\ce{CO2}) model spectra. The expected emission from \ion{O}{1}~1304~\AA\ and \ion{O}{1}~1356~\AA\ from these two sources is then subtracted from the total, as shown in Equations \ref{eq:1} and \ref{eq:2}.}
\label{fig:ei_model}
\end{figure}

An electron impact model for \ce{H2O} and \ce{CO2} derived from \citet{makarov2004kinetic,ajello1971emission} and \citet{shirai2001analytic} is fit to quiescent, background-subtracted data to determine contribution to the \ion{O}{1}~1304~\AA\ and \ion{O}{1} 1356~\AA\ emission features. The expected contribution is subtracted in Equation \ref{eq:1} to prevent an overestimation of the \ce{O2} abundance relative to \ce{H2O}. An example of this model fit is displayed in Figure \ref{fig:ei_model}. The model assumes a \ce{H2O} column density of 10$^{20}$ m$^{-2}$ from VIRTIS measurements \citep{bockelee2016evolution}, 30$\%$ \ce{CO2}/\ce{H2O} relative abundance, 100 eV electron energy, and a Gaussian distribution of photons about the emission feature wavelength, and multiplies the spectrum by a constant until the modeled spectrum resembles the observed. Emission features with threshold energies higher than the average of 15 eV, like the \ion{C}{2} 1335~\AA\ feature, have an additional constant to lower their values. This correction is used to scale for the electron energy distribution at 67P having a significant number of electrons at energies lower than the threshold of these features, but high enough to create \ion{C}{1} 1657~\AA\ or Ly-$\beta$ emission. All electron impact cross sections available from the literature to synthesize the model spectra are taken at 100 eV. 

We use the same method as \citet{feldman2016nature}, which takes advantage of the small \ion{O}{1}~1356~\AA\ cross section for \ce{H2O}. This method requires the assumption that electron impact on \ce{O2} contributing to \ion{O}{1}~1027~\AA\ emission is minimal relative to Ly-$\beta$ emission. The cross section of \ion{O}{1}~1027~\AA\ for \ce{O2} is about an order of magnitude lower than that of Ly-$\beta$ for \ce{H2O}, so this is a reasonable approximation \citep{ajello1985study,makarov2004kinetic}. 

Using the relative cross sections for analysis works under the assumption that the same electron population affects each of the four gases. The model describes the measurements well with minimal adjustment, so this is reasonable assumption to make. This method only yields information on relative abundance, not column density.

\section{Results}\label{results}
The large increase in emission that occurs after 00:00 UTC on October 6 yields the opportunity to explore the composition of the near-nucleus coma, provided there are no simultaneous outbursts of gas from the nucleus. The period during and following the CME impact is of interest due to the correlation of electron density and semi-forbidden \ion{O}{1} 1356 emission multiplet. The increased signal-to-noise ratio for this period allows a qualitative determination of the coma composition during the CME impact for the Alice line of sight. 

\begin{figure}
\centering
\includegraphics[clip,trim=6.5cm 0cm 8.5cm 0cm,width=1.0\linewidth]{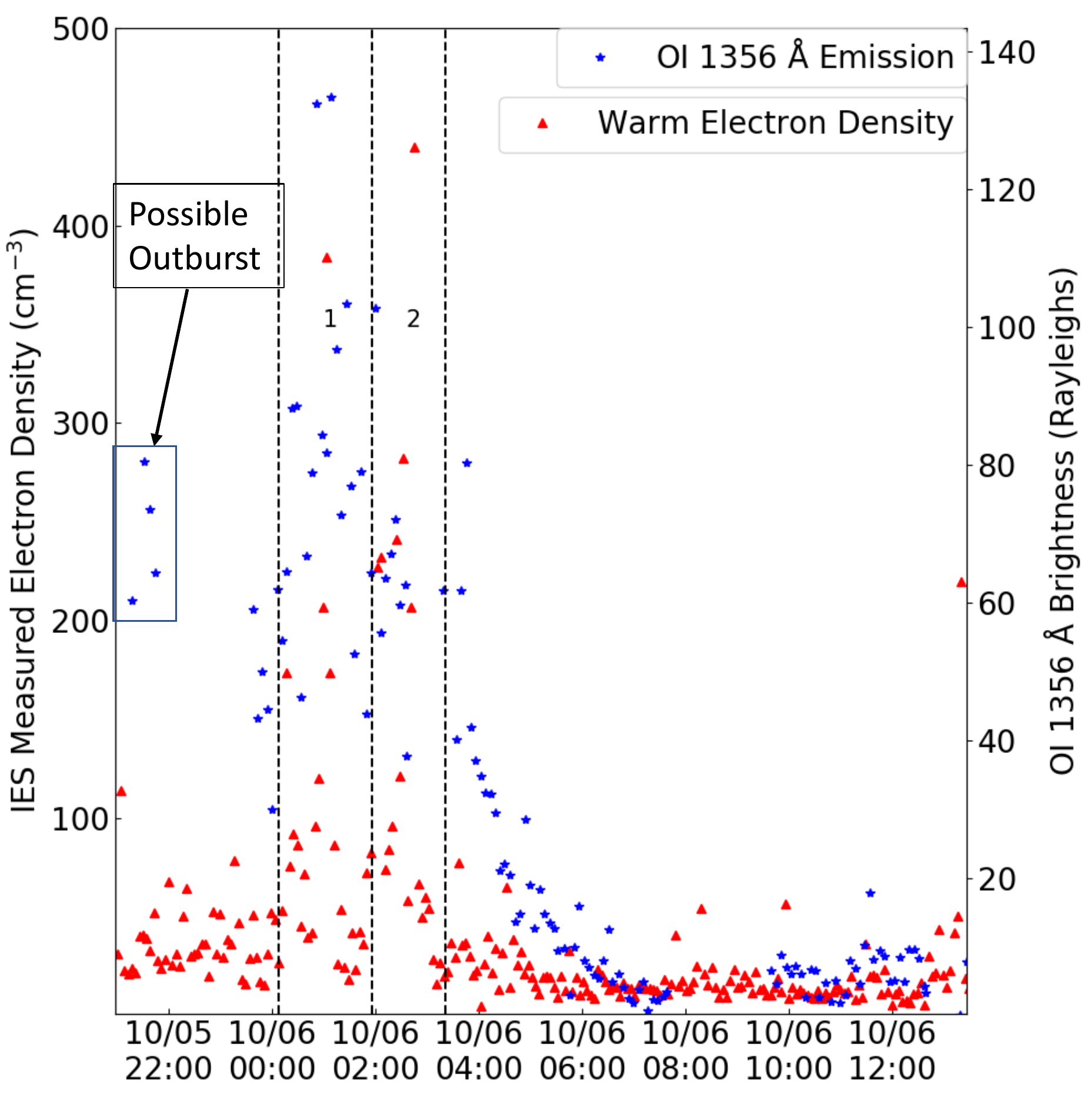}
\caption{Comparison of Alice \ion{O}{1}~1356~\AA\ emission from rows 13--17 (blue stars) with the warm (5--100~eV) electron density (red triangles) as defined by \citet{JGRA:JGRA52830} and \citet{broiles2016statistical} from the IES instrument. The x axis begins at the start of the CME as reported by the RPC magnetometer in \cite{doi:10.1093/mnras/stw2112}. \ion{O}{1} 1356 \AA\ measurements that may indicate a possible outburst are marked. \textcolor{black}{This possible outburst time coincides with  slightly elevated electron fluxes that may be indicative of an outburst as well}. Labels \textbf{1} and \textbf{2} denote the two electron spikes that coincide with spikes in \ion{O}{1}~1356~\AA\ emission. Spectra with an off nadir angle less than 1$^{\circ}$ are used to create the plot.}
\label{fig:OI1356vselectdens}
\end{figure}

\subsection{Concurrent Electron Density Measurements}\label{concurrentobs} 

Using the \ion{O}{1}~1356~\AA\ emission multiplet as a proxy for the electron density near the nucleus, we compare the \ion{O}{1}~1356~\AA\ emission to the measured electron density from the RPC-IES instrument on board the \textit{Rosetta} spacecraft in Figure \ref{fig:OI1356vselectdens}. The electron impact emission of \ion{O}{1}~1356~\AA\ and the electron density both experience an increase starting at 00:00 UTC on October 6. The \ion{O}{1}~1356~\AA\ emission peaks nearly simultaneously with the electron density during the CME arrival and decreases smoothly back to quiescent levels, contrasting the fast drop in electron density measured at the spacecraft after 04:00 UTC (Figure \ref{fig:OI1356vselectdens}). 

The energy distribution of these IES-measured electrons also shows a shift in the energy spectrum. Figure 5 of \cite{doi:10.1093/mnras/stw2112} shows that during the CME impact there is a larger number of electrons with energies $\geq$50 eV than measured in previous days. These observed energies have larger lab-measured dissociative impact cross sections for the relevant molecules, which could explain the increase in emission \citep{makarov2004kinetic,ajello1971dissociative,kanik2003electron,ajello1985study,ajello1971emission_co}. However, the electrons detected by RPC-IES are at the spacecraft, whereas the emission of \ion{O}{1}~1356~\AA\ may come from anywhere along the line of sight within Alice's field of view. This becomes critical during the impact of a coronal mass ejection because, as \cite{doi:10.1093/mnras/stw2112} report, the plasma environment was compressed, \textcolor{black}{allowing solar wind ions to be detected directly by \textit{Rosetta} for the first time since April 2015. }This compression would cause a very different plasma environment at the spacecraft than along the Alice line of sight passing near the nucleus.

The arrival of the CME is characterized by an increase in the electron density and energy, appearance of solar wind ions, and an increase in the magnetic field strength \citep{doi:10.1093/mnras/stw2112}, first occurring at 20:15 UTC October 5. All of these factors were measured by the RPC instruments at \textit{Rosetta}, so the same characteristics may not be applicable across Alice's line of sight. Figure~\ref{fig:OI1356vselectdens} shows that the warm electron population (electrons with energies between 5--100~eV) correlates with \ion{O}{1}~1356~\AA\ emission, with both emission spikes corresponding to maxima of the IES warm electron population. It also appears that the decay in \ion{O}{1}~1356 \AA\ emission with time correlates to the decreasing electron density after the CME passes, around 02:45--03:00 on October 6. The correlation of these two measurements indicates that the CME directly or indirectly increased the electron impact emission of the near-nucleus coma. 

The likelihood of an outburst occurring at the same time as the CME arrival is small, though it cannot be ruled out. The four \ion{O}{1}~1356 \AA\ brightnesses measured an hour after the RPC magnetometer detected the arrival of the CME, but 3 hours before the first steep increases in electron density at the spacecraft, may indicate a gas outburst (Figure \ref{fig:OI1356vselectdens}). If so, this would be an outburst similar to the one detailed in \citet{feldman2016nature} happening just after the CME arrival. A first order calculation of the probability of an outburst overlapping with the CME using the outburst frequency of 0.78 outbursts/day from \citet{vincent2016summer} suggests a 2\% chance of this particular case. These observations may also be an indication of a more rapid change in the near-nucleus electron environment following the CME impact, but without simultaneous electron measurements at both locations there is unfortunately no way to disentangle the two possibilities.

\begin{table*}[t!]
\begin{center}
\begin{tabular}{ c c c c c c c c }
\hline
Observation & Time (UTC) & d$_{comet}$ (km) & Scan Rate ($^{\circ}$/s) & Off-Nadir Angle ($^{\circ}$) &$\frac{OI~1304}{OI~1356}$ & $\frac{CI~1657}{HI~Ly-\beta}$ & $\frac{HI~Ly-\beta}{OI~1304}$ \\[.5ex]
\hline
October 5 Pre-CME & 00:08--00:38 & 875 & 6.36E-5 & 0.54 &2.8$\pm$0.6 & 0.64$\pm$0.05 & 0.96$\pm$0.03 \\[.5ex]
October 6 CME Spike 1 & 00:46--00:57 & 763 & 8.29E-5 & 0.45 &0.96$\pm$0.7 & 0.73$\pm$0.1 &0.86$\pm$0.01 \\[.5ex]
October 6 CME Spike 2 & 02:29--02:48 & 756 & 8.64E-5 & 0.44 &1.2$\pm$0.1 & 0.47$\pm$0.08 & 1.20$\pm$0.07 \\[.5ex]
October 6 Post-CME & 06:42--06:52 & 737 & 1.38E-4 & 0.52 &4.3$\pm$1.4 & 0.33$\pm$0.2 & 1.7$\pm$0.6 \\[.5ex]
 \hline
\end{tabular}
\caption{Observed emission line ratios for four distinct periods described in Sections 3 and 4. Pre-CME values are taken from three exposures made two comet rotations earlier with similar, though not identical, pointing to the spectra plotted in Figure~\ref{fig:1005_spectra}a. All values are taken from rows 13-17. CME spike values correspond to the three exposures closest to the maximum spectrum for each spike, both shown in Figure~\ref{fig:CME_Spectra}. The post-CME values are calculated from three exposures gathered just before the final gap in data at 07:37 UTC on October 6. \textcolor{black}{This period is used as the quiescent subtraction due to the identical pointing.}}
\label{table:2}
\end{center}
\end{table*}
\subsection{Spectra}\label{spectra}
The most likely cause of the spikes in emission is the CME, whether through direct impact of CME electrons or higher order interactions, such as magnetic reconnection events or ionization of neutrals by CME energetic particles. In either case the significant presence of the semi-forbidden \ion{O}{1}~1356~\AA\ line is an indication that the emission spike has a large electron impact component; as a spin-forbidden transition it can only occur from electron impact and not resonance fluorescence. Providing there was not a gas outburst from the comet at a coinciding time this data would provide a sampling of the near-nucleus coma. If this is the case, a brief comparison of the line ratios during the CME to Table \ref{table:1} shows that the portion of the coma observed would be in reasonable agreement with a mixture of H$_{2}$O and O$_{2}$ plus a small component of CO$_{2}$. This mixture would produce spectral features similar to the outburst composition of \ce{O2}/\ce{H2O} $\geq$ 0.5 and a \ion{C}{1} 1657 \AA\ emission with an unclear origin found by \citet{feldman2016nature}. Though sulfur and sulfur-bearing compounds have been observed at 67P, the observed sulfur multiplet emission does not correspond to electron impact on \ce{SO2} \citep{calmonte2016sulphur,JGRA:JGRA16941}. By subtracting the quiescent period spectra from the spectra taken during the CME we can attempt to identify the composition in the coma at the time of the CME and examine the effect the solar event had on the coma. The line ratios of four periods of specific interest are summarized in Table \ref{table:2} and analyzed here.

\subsubsection{Pre CME Emission}\label{precmeemission}
The first period of interest covers four observations made between 00:08 and 00:38 UTC, during which the Alice slit intersects the nucleus of 67P. The emission is consistent with the October 5 00:49 UTC spectrum plotted in Figure~\ref{fig:CME_Spectra}. This time period is characterized by low levels of emission of the Ly-$\beta$, \ion{O}{1}~1304~\AA, \ion{O}{1}~1356~\AA, and \ion{C}{1}~1657~\AA\ multiplets, most likely indicative of the pre-CME coma environment. The emission multiplet ratios from Table \ref{table:2} for the rows closest to the nucleus, and thus most affected by electron impact excitation due to the line of sight integration and higher column density, indicate a \ce{H2O}-dominant coma with carbon compounds contributing to the \ion{C}{1} 1657~\AA\ multiplet, but with no obvious \ce{O2} signature. There is a nearly 24-hour time difference between this period of interest and the first emission spike during the CME impact. This time difference opens the possibility that the composition of the coma seen by Alice two comet rotations prior to the CME impact was different from the coma composition observed during the CME impact. 

\subsubsection{First Emission Spike}\label{firstspike}
Emission from the coma reaches the first spike on October 6th at 00:52 UTC, just over an hour after the initial CME impact \citep{doi:10.1093/mnras/stw2112}. Spectra taken between 00:46 and 00:57 on October 6 are used to characterize this spike. This period corresponds to the maximum density of solar wind ions measured by the ICA instrument during the CME (see figure~4b of \citealt{doi:10.1093/mnras/stw2112}). Table \ref{table:2} and Figure~\ref{fig:151005_151007_Ratio_Coma} show that the line ratios that occur during the CME are not similar to what was observed during two cometary rotations earlier; the \ion{O}{1}~1304/1356~\AA\ ratio has dropped to $\approx$1, indicating the presence of \ce{O2} \citep{feldman2016nature,kanik2003electron}. The \ion{C}{1}~1657~\AA/Ly-$\beta$ ratio for this period increases from 0.64 to 0.73, and the Ly-$\beta$/\ion{O}{1}~1304~\AA\ ratio decreases from 0.96 to 0.86. Following this spike all emission lines experience a decline until the second spike occurs. \textcolor{black}{Additionally, this period is missing CO emission from electron impact on \ce{CO2}, which has maximum cross sections between 20-40 eV  \citep{ajello1971emission}. This suggests that the dominant electrons are in the 100 eV range, where cross sections are maximized for \ce{O2}, \ce{H2O}, and \ce{CO2} carbon and oxygen emission.}

\subsubsection{Second Emission Spike}\label{secondspike}
The second spike occurs approximately 1.5 hours after the first, with the maximum reached at 02:34 UTC during a short 46-second exposure. Due to the short exposure time the signal-to-noise ratio is lower than the surrounding exposures at 02:29 and 02:48 UTC. Measurements from the IES instrument show the highest density and energy of electrons occur during this time period (Figure~\ref{fig:OI1356vselectdens} of this paper and Figure~4c of \citealt{doi:10.1093/mnras/stw2112}). Again, we see that the emission increases for all of the largest multiplets, but unlike the first spike Ly-$\beta$ experiences the largest relative increase. The \ion{C}{1} 1657~\AA/Ly-$\beta$ ratio dropped further from the first spike, down to 0.47. Similarly, the Ly-$\beta$/\ion{O}{1}~1304~\AA\ ratio increases up to 1.2, the result of a stronger Ly-$\beta$ presence in the second spike and a decreased presence of \ion{O}{1}~1304~\AA\ and \ion{O}{1}~1356~\AA\ emission (see Figure~\ref{fig:Smoothed_October}). The \ion{O}{1}~1304/1356~\AA\ ratio rose slightly to 1.2, indicating a lower abundance of \ce{O2} in the coma. The increases seen with Ly-$\beta$ also suggest a change to the \ce{O2}/\ce{H2O} relative abundance. \textcolor{black}{Individual spectra near this spike show some evidence of electron impact on \ce{CO2} producing CO emission (Fig \ref{fig:CME_Spectra}) in the 1400 to 1500~\AA\ region. The CO emission suggests that the incident electron population has a cooler population with energies nearer to the CO emission from dissociative electron impact on \ce{CO2} \citep{ajello1971emission}.} Alice observations cease before a decrease in the emission spike is observed, leaving the possibility that the increase in emissions continued. 

\subsubsection{Post CME Emission}\label{postcmeemission}
\begin{figure}
\centering
\includegraphics[clip,trim=0.5cm 0cm 1cm 0cm,width=1.0\linewidth]{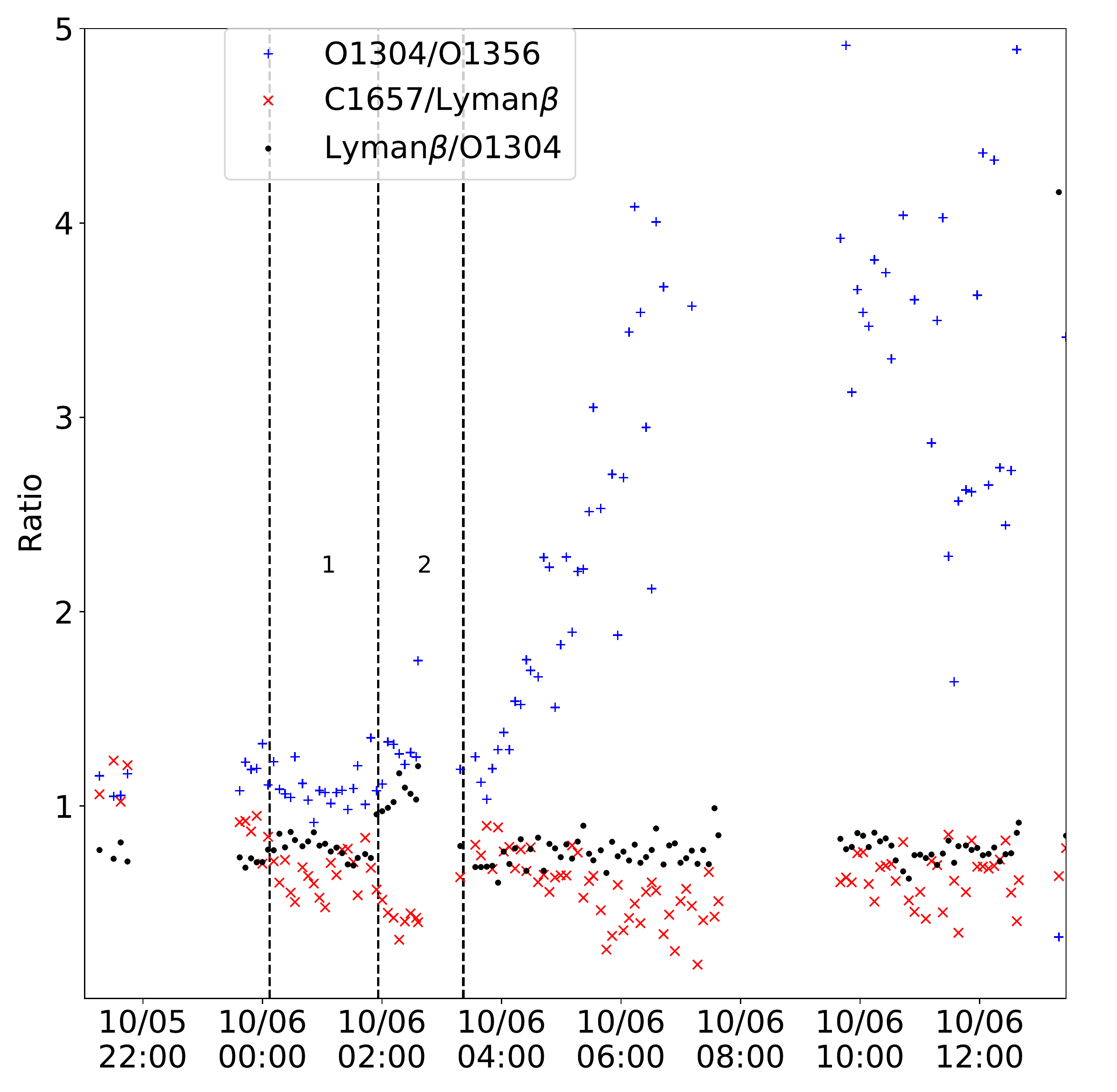}
\caption{Emission line ratios as a function of time from rows 13-17. Labels \textbf{1} and \textbf{2} mark the same boundaries for electron spikes as in Figure \ref{fig:OI1356vselectdens}. The x-axis starts at the time of the first RPC detection of the CME arrival at 20:15 UTC. Of particular interest is the \ion{O}{1}~1304/1356 \AA\ ratio near 1 during the impact, the drop in \ion{C}{1} 1657/Ly-$\beta$ \AA, and the increase to Ly-$\beta$/\ion{O}{1}~1304 \AA\ and \ion{O}{1}~1304/1356 \AA\ ratios during the secondary outburst.}
\label{fig:151005_151007_Ratio_Coma}
\end{figure}

As observations resumed again at 03:15 UTC all emissions experienced a steep decrease down to a background level (Figure~\ref{fig:Smoothed_October}). This smooth decline stands in contrast to the sharp drop seen in the electron density by IES at the spacecraft (Figure~\ref{fig:OI1356vselectdens}), suggesting a difference between the near-nucleus and spacecraft-measured electron populations. Using the \ion{O}{1}~1356~\AA\ multiplet as a proxy for electron impact emission shows that the contribution of electron impact to the emissions almost entirely disappears. The \ion{O}{1}~1304/1356 ratio rose to 5.3 due to the decrease in the electron impact emission, and continued to rise after this time period (Figure~\ref{fig:151005_151007_Ratio_Coma}). The changes in Lyman-$\beta$ and the \ion{C}{1} 1657~\AA\ multiplet seen in the second emission spike continue, with the \ion{C}{1} 1657~\AA/Ly-$\beta$ ratio measured near 0.4--0.8 from the resumption of observations at 03:15 UTC onward. For the post-CME time period described here the value was 0.35. Compare this trend to the period just prior to the second emission spike in Figure~\ref{fig:151005_151007_Ratio_Coma}, which shows a steady increase to the ratio prior to the end of observations.  The lack of electron impact emission for this time period prevents the accurate use of Table \ref{table:1} for analyzing composition. 

\section{Discussion}\label{discussion}
 The emissions spikes observed on October 6 present several problems for decisive analysis. Due to the timing of the emission compared to the CME impact it seems most likely that the emission is driven by changes to the electron environment, though we cannot rule out that an outburst occurred at the same time. The four significant detections of \ion{O}{1}~1356~\AA\ around October 5 21:25 UTC, approximately 3 hours before the substantial increase to the RPC-IES measured electron density (Figure \ref{fig:OI1356vselectdens}), may be a sign of an outburst similar to that reported in \cite{feldman2016nature}. Here we attempt to distinguish between possible scenarios that could have increased electron energy and/or density and how they compare to observations.

\subsection{Electron Impact Emission}\label{EIimpact_discussion}
The detection of the semi-forbidden \ion{O}{1}~1356~\AA\ emission line for the duration of the CME as measured by IES supports the hypothesis that the change in emissions during this period was the result of increased electron impact on the coma of 67P (Figure~\ref{fig:OI1356vselectdens}). The 1:1 ratio of \ion{O}{1}~1304~\AA\ and \ion{O}{1}~1356~\AA\ in Figure~\ref{fig:151005_151007_Ratio_Coma} is a prime indicator of the electron environment's effect on the coma, since the only way for these to reach equal levels is as a result of electron impact \citep{kanik2003electron,ajello1971dissociative,ajello1971emission}. Due to the unique circumstances surrounding these observations we would like to address several hypotheses for how the electron environment could have become more favorable for electron impact emission during the CME impact. 
\subsubsection{Introduction of CME Electrons}\label{CME_electrons}
The simplest case is that the increase in electron impact could result from the introduction of CME electrons to the near-nucleus environment. The RPC instruments on board \textit{Rosetta} observed and reported on this electron population in \citet{doi:10.1093/mnras/stw2112}, which was rich with electrons in the 10--200~eV energy range. If this population of electrons penetrated into the near-nucleus environment, the energies would be ideal for maximizing emission from electron impact based on lab-determined cross sections. Because the RPC measurements are taken \textit{in situ} and Alice results represent a line-of-sight integration, assumptions must be made about the electron and gas density along the line of sight in order to properly determine this effect's contribution. However, this hypothesis does not explain the difference in slope between the observed \ion{O}{1}~1356~\AA\ emission and the \textit{in situ} electron measurements, which would be expected to match exactly if the CME electrons were the main contributor due to the short lifetime of the excited state, and under the assumption of uniformity for the CME electron density on the scale of the \textit{Rosetta}-comet distance. This mismatch between slopes, especially in the period following 04:00 UTC, is clearly seen in Figure~\ref{fig:OI1356vselectdens}. 

\begin{table*}[t!]
\begin{center}
	\begin{tabular}{c c c c c}
		\hline
		Observation & \multicolumn2c{$\frac{\ce{O2}}{\ce{H2O}}$ from Equation 1} & \multicolumn2c{$\frac{\ce{O2}}{\ce{H2O}}$ from Equation 2}\\ 
		\cline{2-5}
		 & Rows 13-17 & Rows 18-22
		 & Rows 13-17 & Rows 18-22 \\
		\hline
		October 6 CME Spike 1 & 0.14$\pm$0.01 & 0.13$\pm$0.02 & 0.14$\pm$0.01 & 0.15$\pm$0.01 \\ 
		October 6 CME Spike 2 & 0.10$\pm$0.01 & 0.06$\pm$0.01 & 0.08$\pm$0.01 & 0.05$\pm$0.02 \\
		\hline
	\end{tabular}
\caption{\textcolor{black}{Calculated \ce{O2}/\ce{H2O} abundances from the emission spikes described in Sections \ref{observations}/\ref{results} and plotted in Figure~\ref{fig:CME_Spectra}.}}
\label{table:3}
\end{center}
\end{table*}

\subsubsection{Compression of the Diamagnetic Cavity}\label{plasma_compression}
The CME impact onto the comet likely compressed the plasma environment of the coma, \textcolor{black}{ allowing solar wind ions to penetrate closer to the nucleus for the first time since March 2015 \citep{doi:10.1093/mnras/stw2112}. The most important aspect of the CME's effect on the environment for dissociative electron impact emission is the compression of the diamagnetic cavity, within which there is no magnetic field. At 67P the region inside the diamagnetic cavity was determined to be somewhat depleted of electrons between 150-200 eV and substantially depleted of electrons around 100 eV \citep{nemeth2016charged}, making the cavity less favorable for electron impact emission. Furthermore, the electron gyroradius is small compared to other length scales in the plasma of the coma environment, preventing electrons from the extended coma and solar wind from passing into the cavity. This would imply that the electron population best suited for dissociative electron impact excitation exists just outside the cavity, where electrons have the highest density and energy distribution and the neutral number density is highest. }

\textcolor{black}{The diamagnetic cavity radius was first calculated using a balance of the Lorentz and neutral friction force \citep{cravens1987theory,ip1987formation}, but we can now bolster this with observational constraints from RPC measurements. At 67P the diamagnetic cavity was found to extend farther from the nucleus then expected \citep{goetz2016first,goetz2016structure}, and can be calculated using: 
\begin{equation}
r_{c} = \left(\frac{B(r)^{2}}{c^{2}Q^{3/2}}+\frac{1}{r^{2}}\right)^{-1/2},
\end{equation}
where \textit{B(r)} is the magnetic field measured at radius \textit{r}, \textit{c} is the constant 7.08$\times$10$^{-18}$ km nT s$^{3/4}$, and \textit{Q} is the production rate of the comet \citep{timar2017modelling}.}

\textcolor{black}{If the production rate of the comet is assumed to be constant during the passage of the CME the radius of the diamagnetic cavity can be calculated using the magnetometer measurements stated in \citet{doi:10.1093/mnras/stw2112} and spacecraft-comet radius. In the initial conditions, with a measured magnetic field magnitude of 40 nT, a water production rate of 7$\times$10$^{27}$ s$^{-1}$ from \citet{hansen2016evolution}, and a spacecraft-comet distance radius of 876 km this corresponds to a cavity extending 134 km from the nucleus. For magnetic field magnitudes of 60 and 100 nT and spacecraft-comet radii of 766 and 756 km for emission spikes 1 and 2 this corresponds to radii of 85 and 54 km, respectively. }

\textcolor{black}{These two different cavity radii probe regions of the coma with approximately 2.5 and 6 times the number density of neutrals at the original cavity radii, if a Haser model of neutral distribution is assumed. When the higher density and energy electron population of the CME are coupled with the higher neutral density of the inner coma regions, the area most favored for dissociative electron impact emission is a shell just outside of the diamagnetic cavity boundary. Taking into account the factor of 2.5-6 increase in neutral density, the factor of 6-10 higher electron flux from the CME, and the factor of 2-3 higher average electron energy we see the electron impact emission could be expected to increase between a factor of 10-20 over the quiescent values, depending on the emission spike and time of RPC observations \citep{edberg2016solar}. This increase is similar in magnitude to the spikes show in Figure \ref{fig:Smoothed_October}.}

\textcolor{black}{The orientation of the Alice slit during this time period, parallel to the Sun-comet line and within a degree of the nucleus in the middle of the slit, means that even when the diamagnetic cavity was most compressed in emission spike 2 the line of sight for all rows still passed through the outer coma, through the diamagnetic cavity, and back into the outer coma on the other side. Throughout the CME impact the middle and upper rows of the detector would have captured these two boundary regions of the cavity, providing additional continuity to the observation pointing and geometry. This should allow the first order comparison done above to hold for these situations, but there are some caveats. The structure of the diamagnetic cavity has been shown to have sinusoidal heterogeneities in the structure \citep{goetz2016structure,henri2017diamagnetic}, and it is possible that there were changes to the structure and radius of the diamagnetic cavity due to the CME.}

The subsequent expansion of the \textcolor{black}{diamagnetic cavity} following the passage of the CME and a subsequent decrease in magnetic field magnitude could explain the smooth decline in \ion{O}{1}~1356~\AA\ emission. More simulation work is required to further explore this possibility, specifically with magnetohydrodynamic and hybrid modeling to properly constrain the behavior of the magnetic field lines near the nucleus in these direct CME impact cases. 

\subsubsection{Lower Hybrid Waves}\label{lower_hybrid_waves}
Additional plasma physics could also contribute to the increase in dissociative electron impact emission. The lower hybrid waves observed in the plasma environment of 67P by the RPC instruments may have played a role. These waves, which are the result of lower hybrid drift instabilities in the plasma, were observed by the Langmuir probe (LAP) on board \textit{Rosetta} in October and November of 2015, approximately the same time post-perihelion as the CME impact \citep{GRL:GRL55530}. The ion and electron gradients that drive the instabilities creating lower hybrid waves are heavily influenced by interactions with the solar wind, so an energetic event like the CME could have drastically amplified the waves observed just a few weeks later by LAP. Lower hybrid waves are capable of heating the thermal electron population (5--10~eV) to energies above the threshold energy for electron impact emission (15--20~eV) \citep{GRL:GRL55530,andre2017lower}. A boost to this super-threshold population from increased lower hybrid waves during the CME impact could explain the decoupling between the electron density and dissociative electron impact emission after the second emission spike shown in Figure \ref{fig:OI1356vselectdens}, where it is clear that there is a divergence between the electrons measured by RPC and the impacting electron population along the Alice line of site. 

\subsection{$O_{2}$ in the Coma}\label{coma_O2}
The strong appearance of \ion{O}{1}~1304~\AA\ and \ion{O}{1}~1356~\AA\ emission in the spectra taken by Alice indicates that there is a substantial amount of \ce{O2} present in the coma of 67P at the time of the CME impact, or introduced to the near-nucleus coma from an outburst. \ce{O2}/\ce{H2O} abundances calculated using Equations \ref{eq:1} and \ref{eq:2} on a sample of \textcolor{black}{three} spectra centered on the maxima of emission spikes 1 and 2 are shown in Table \ref{table:3}. The lower rows of the slit are not used in analysis due to the decreased dissociative electron impact emission observed there. The first emission spike has an average \ce{O2}/\ce{H2O} ratio of 0.14. The second emission spike has an average value of 0.08, just over half that of the first. 

These calculated values show an \ce{O2}/\ce{H2O} relative abundance that ranges from two to five times that of the \ce{O2}/\ce{H2O} ratio of 0.038 found by \citet{bieler2015abundant} and below that of the \ce{O2}/\ce{H2O} ratio of \textcolor{black}{$\sim$ 0.22} found by \citet{feldman2016nature} for several outbursts in 2015. This level of \ce{O2} in the coma is not unique, however. Stellar appulse observations taken with Alice in 2015 show a range of 0.1 to 0.6 for \ce{O2}/\ce{H2O} \citep{keeney2017h2o}. The drop in the relative abundance between the first and second emission spikes suggests a change to the coma composition in the hour and half between them, which may or may not be related to the CME. All cases suggest that the presence of \ce{O2} at 67P is substantial, which requires mechanisms for trapping the highly volatile \ce{O2} into ice and/or for forming \ce{O2} through chemical pathways \citep{mousis2016origin,taquet2016primordial,dulieu2017production}. 

\section{Summary}\label{summary}
Based on the comparison between the IES measured electron densities, cross sections of water, carbon dioxide, and molecular oxygen and the observed line ratios for FUV spectra taken during the CME impact on 2015 October 5--6, we believe that substantial electron impact dissociation took place. Although the exact source of the increased emission cannot be specifically stated, the timing of the emission spikes matches the arrival of the electrons attributed to the CME. The unique electron environment allowed Alice to observe the near-nucleus coma environment in a way that had previously only affected a region within tens of km of the surface \citep{feldman2015measurements}. Two emission spikes correlate to IES measurements of increased electron density, magnetometer measurements of increased magnetic field magnitudes, and have two different \ce{O2}/\ce{H2O} ratios, indicating \textcolor{black}{ change to the region affected by electron impact emission in the 90 minutes separating the spikes}. The emission along the Alice line of sight decays over the next several hours back to the quiescent level following a steep drop in the warm electron density as measured by IES at the \textit{Rosetta} spacecraft. The near nucleus environment experienced profound changes during the CME impact that resulted in the dominance of electron impact emission for the duration. This period of increased emission was used to calculate the \ce{O2}/\ce{H2O} abundance ratio, ranging from 0.06--0.16. This research supports the results of \cite{bieler2015abundant}, who found that the levels of molecular oxygen are high enough to no longer fit current cometary formation models, and that the process that creates these reservoirs of molecular oxygen in the comet is still unknown. However, the \ce{O2}/\ce{H2O} ratio in this event was several times higher than the result of \citet{bieler2015abundant}. The \ce{O2}/\ce{H2O} ratio values found by this work are lower than the ratio found by \citet{feldman2016nature} and agree with low impact parameter values from stellar appulse observations \citep{keeney2017h2o}. 

\acknowledgements
The research presented here was made possible by the 
ESA/NASA \textit{Rosetta} mission with contributions from ESA member states and NASA. The Alice team would like to acknowledge the support of NASA's Jet Propulsion Laboratory, specifically through contract 1336850 to the Southwest Research Institute. Work at University of Oslo was supported by the Research Council of Norway grant No. 240000. We also want to thank our reviewer for their insightful feedback and edits. We would like to acknowledge ISSI for offering us the opportunity to have very valuable discussions on this topic as part of the International Team 'Plasma Environment of comet 67P after Rosetta (402)'.
\bibliographystyle{aasjournal} 
\bibliography{uv_cme_impact}

\begin{thebibliography}{}
\expandafter\ifx\csname natexlab\endcsname\relax\def\natexlab#1{#1}\fi
\providecommand{\url}[1]{\href{#1}{#1}}

\bibitem[{Ajello \& Franklin(1985)}]{ajello1985study}
Ajello, J., \& Franklin, B. 1985, \jcp, 82, 2519

\bibitem[{Ajello(1971{\natexlab{a}})}]{ajello1971dissociative}
Ajello, J.~M. 1971{\natexlab{a}}, \jcp, 55, 3156

\bibitem[{Ajello(1971{\natexlab{b}})}]{ajello1971emission}
---. 1971{\natexlab{b}}, \jcp, 55, 3169

\bibitem[{Ajello(1971{\natexlab{c}})}]{ajello1971emission_co}
---. 1971{\natexlab{c}}, \jcp, 55, 3158

\bibitem[{Andr{\'e} {et~al.}(2017)Andr{\'e}, Odelstad, Graham, Eriksson,
  Karlsson, Stenberg~Wieser, Vigren, Norgren, Johansson, Henri,
  {et~al.}}]{andre2017lower}
Andr{\'e}, M., Odelstad, E., Graham, D., {et~al.} 2017, Monthly Notices of the
  Royal Astronomical Society, 469, S29

\bibitem[{Bieler {et~al.}(2015)Bieler, Altwegg, Balsiger, Bar-Nun, Berthelier,
  Bochsler, Briois, Calmonte, Combi, De~Keyser, {et~al.}}]{bieler2015abundant}
Bieler, A., Altwegg, K., Balsiger, H., {et~al.} 2015, \nat, 526, 678

\bibitem[{Bockel{\'e}e-Morvan {et~al.}(2016)Bockel{\'e}e-Morvan, Crovisier,
  Erard, Capaccioni, Leyrat, Filacchione, Drossart, Encrenaz, Biver,
  de~Sanctis, {et~al.}}]{bockelee2016evolution}
Bockel{\'e}e-Morvan, D., Crovisier, J., Erard, S., {et~al.} 2016, \mnras, 462,
  S170

\bibitem[{Broiles {et~al.}(2016{\natexlab{a}})Broiles, Livadiotis, Burch, Chae,
  Clark, Cravens, Davidson, Eriksson, Frahm, Fuselier, Goldstein, Goldstein,
  Henri, Madanian, Mandt, Mokashi, Pollock, Rahmati, Samara, \&
  Schwartz}]{JGRA:JGRA52830}
Broiles, T.~W., Livadiotis, G., Burch, J.~L., {et~al.} 2016{\natexlab{a}},
  \jgr: Space Physics, 121, 7407, 2016JA022972.
\newblock \url{http://dx.doi.org/10.1002/2016JA022972}

\bibitem[{Broiles {et~al.}(2016{\natexlab{b}})Broiles, Burch, Chae, Clark,
  Cravens, Eriksson, Fuselier, Frahm, Gasc, Goldstein,
  {et~al.}}]{broiles2016statistical}
Broiles, T.~W., Burch, J., Chae, K., {et~al.} 2016{\natexlab{b}}, \mnras, 462,
  S312

\bibitem[{Burch {et~al.}(2007)Burch, Goldstein, Cravens, Gibson, Lundin,
  Pollock, Winningham, \& Young}]{burch2007rpc}
Burch, J., Goldstein, R., Cravens, T., {et~al.} 2007, \ssr, 128, 697

\bibitem[{Calmonte {et~al.}(2016)Calmonte, Altwegg, Balsiger, Berthelier,
  Bieler, Cessateur, Dhooghe, Van~Dishoeck, Fiethe, Fuselier,
  {et~al.}}]{calmonte2016sulphur}
Calmonte, U., Altwegg, K., Balsiger, H., {et~al.} 2016, \mnras, 462, S253

\bibitem[{Chaufray {et~al.}(2017)Chaufray, Bockel{\'e}e-Morvan, Bertaux, Erard,
  Feldman, Capaccioni, Schindhelm, Leyrat, Parker, Filacchione,
  {et~al.}}]{chaufray2017rosetta}
Chaufray, J.-Y., Bockel{\'e}e-Morvan, D., Bertaux, J.-L., {et~al.} 2017,
  Monthly Notices of the Royal Astronomical Society, 469, S416

\bibitem[{Clark {et~al.}(2015)Clark, Broiles, Burch, Collinson, Cravens, Frahm,
  Goldstein, Goldstein, Mandt, Mokashi, {et~al.}}]{clark2015suprathermal}
Clark, G., Broiles, T., Burch, J., {et~al.} 2015, \aap, 583, A24

\bibitem[{Coradini {et~al.}(2007)Coradini, Capaccioni, Drossart, Arnold,
  Ammannito, Angrilli, Barucci, Bellucci, Benkhoff, Bianchini,
  {et~al.}}]{coradini2007virtis}
Coradini, A., Capaccioni, F., Drossart, P., {et~al.} 2007, Space Science
  Reviews, 128, 529

\bibitem[{Cravens(1987)}]{cravens1987theory}
Cravens, T.~E. 1987, Advances in space research, 7, 147

\bibitem[{Dulieu {et~al.}(2017)Dulieu, Minissale, \&
  Bockel{\'e}e-Morvan}]{dulieu2017production}
Dulieu, F., Minissale, M., \& Bockel{\'e}e-Morvan, D. 2017, \aap, 597, A56

\bibitem[{Edberg {et~al.}(2016{\natexlab{a}})Edberg, Eriksson, Odelstad,
  Vigren, Andrews, Johansson, Burch, Carr, Cupido, Glassmeier,
  {et~al.}}]{edberg2016solar}
Edberg, N.~J., Eriksson, A.~I., Odelstad, E., {et~al.} 2016{\natexlab{a}},
  \jgr: Space Physics, 121, 949

\bibitem[{Edberg {et~al.}(2016{\natexlab{b}})Edberg, Alho, Andr{\'e}, Andrews,
  Behar, Burch, Carr, Cupido, Engelhardt, Eriksson, Glassmeier, Goetz,
  Goldstein, Henri, Johansson, Koenders, Mandt, Möstl, Nilsson, Odelstad,
  Richter, Simon~Wedlund, Stenberg~Wieser, Szego, Vigren, \&
  Volwerk}]{doi:10.1093/mnras/stw2112}
Edberg, N. J.~T., Alho, M., Andr{\'e}, M., {et~al.} 2016{\natexlab{b}}, \mnras,
  462, S45.
\newblock \url{+ http://dx.doi.org/10.1093/mnras/stw2112}

\bibitem[{Feldman {et~al.}(2015)Feldman, A'Hearn, Bertaux, Feaga, Parker,
  Schindhelm, Steffl, Stern, Weaver, Sierks,
  {et~al.}}]{feldman2015measurements}
Feldman, P.~D., A'Hearn, M.~F., Bertaux, J.-L., {et~al.} 2015, \aap, 583, A8

\bibitem[{Feldman {et~al.}(2016)Feldman, A'Hearn, Feaga, Bertaux, Noonan,
  Parker, Schindhelm, Steffl, Stern, \& Weaver}]{feldman2016nature}
Feldman, P.~D., A'Hearn, M.~F., Feaga, L.~M., {et~al.} 2016, \apj Letters, 825,
  L8

\bibitem[{Feldman {et~al.}(2018)Feldman, A’Hearn, Bertaux, Feaga, Keeney,
  Knight, Noonan, Parker, Schindhelm, Steffl, {et~al.}}]{feldman2018fuv}
Feldman, P.~D., A’Hearn, M.~F., Bertaux, J.-L., {et~al.} 2018, The
  Astronomical Journal, 155, 9

\bibitem[{Fougere {et~al.}(2016)Fougere, Altwegg, Berthelier, Bieler,
  Bockelee-Morvan, Calmonte, Capaccioni, Combi, De~Keyser, Debout,
  {et~al.}}]{fougere2016three}
Fougere, N., Altwegg, K., Berthelier, J.-J., {et~al.} 2016, \aap, 588, A134

\bibitem[{Goetz {et~al.}(2016{\natexlab{a}})Goetz, Koenders, Richter, Altwegg,
  Burch, Carr, Cupido, Eriksson, G{\"u}ttler, Henri, {et~al.}}]{goetz2016first}
Goetz, C., Koenders, C., Richter, I., {et~al.} 2016{\natexlab{a}}, Astronomy \&
  Astrophysics, 588, A24

\bibitem[{Goetz {et~al.}(2016{\natexlab{b}})Goetz, Koenders, Hansen, Burch,
  Carr, Eriksson, Fr{\"u}hauff, G{\"u}ttler, Henri, Nilsson,
  {et~al.}}]{goetz2016structure}
Goetz, C., Koenders, C., Hansen, K., {et~al.} 2016{\natexlab{b}}, Monthly
  Notices of the Royal Astronomical Society, 462, S459

\bibitem[{Gray {et~al.}(2014)Gray, Chanover, Slanger, \&
  Molaverdikhani}]{gray2014effect}
Gray, C., Chanover, N., Slanger, T., \& Molaverdikhani, K. 2014, Icarus, 233,
  342

\bibitem[{Haider {et~al.}(2009)Haider, Abdu, Batista, Sobral, Kallio, Maguire,
  \& Verigin}]{GRL:GRL26046}
Haider, S.~A., Abdu, M.~A., Batista, I.~S., {et~al.} 2009, \grl, 36, n/a,
  l13104.
\newblock \url{http://dx.doi.org/10.1029/2009GL038694}

\bibitem[{Hansen {et~al.}(2016)Hansen, Altwegg, Berthelier, Bieler, Biver,
  Bockel{\'e}e-Morvan, Calmonte, Capaccioni, Combi, De~Keyser,
  {et~al.}}]{hansen2016evolution}
Hansen, K.~C., Altwegg, K., Berthelier, J.-J., {et~al.} 2016, \mnras, stw2413

\bibitem[{Henri {et~al.}(2017)Henri, Valli{\`e}res, Hajra, Goetz, Richter,
  Glassmeier, Galand, Rubin, Eriksson, Nemeth, {et~al.}}]{henri2017diamagnetic}
Henri, P., Valli{\`e}res, X., Hajra, R., {et~al.} 2017, Monthly Notices of the
  Royal Astronomical Society, 469, S372

\bibitem[{Ip \& Axford(1987)}]{ip1987formation}
Ip, W.-H., \& Axford, W. 1987, Nature, 325, 418

\bibitem[{Kanik {et~al.}(2003)Kanik, Noren, Makarov, Vattipalle, Ajello, \&
  Shemansky}]{kanik2003electron}
Kanik, I., Noren, C., Makarov, O., {et~al.} 2003, \jgr: Planets, 108

\bibitem[{Karlsson {et~al.}(2017)Karlsson, Eriksson, Odelstad, André, Dickeli,
  Kullen, Lindqvist, Nilsson, \& Richter}]{GRL:GRL55530}
Karlsson, T., Eriksson, A.~I., Odelstad, E., {et~al.} 2017, \grl, 44, 1641,
  2016GL072419.
\newblock \url{http://dx.doi.org/10.1002/2016GL072419}

\bibitem[{Keeney {et~al.}(2017)Keeney, Stern, A’hearn, Bertaux, Feaga,
  Feldman, Medina, Parker, Pineau, Schindhelm, {et~al.}}]{keeney2017h2o}
Keeney, B.~A., Stern, S.~A., A’hearn, M.~F., {et~al.} 2017, Monthly Notices
  of the Royal Astronomical Society, 469, S158

\bibitem[{Makarov {et~al.}(2004)Makarov, Ajello, Vattipalle, Kanik, Festou, \&
  Bhardwaj}]{makarov2004kinetic}
Makarov, O.~P., Ajello, J.~M., Vattipalle, P., {et~al.} 2004, \jgr: Space
  Physics, 109

\bibitem[{Mall {et~al.}(2016)Mall, Altwegg, Balsiger, Bar-Nun, Berthelier,
  Bieler, Bochsler, Briois, Calmonte, Combi, {et~al.}}]{mall2016high}
Mall, U., Altwegg, K., Balsiger, H., {et~al.} 2016, The Astrophysical Journal,
  819, 126

\bibitem[{Mousis {et~al.}(2016)Mousis, Ronnet, Brugger, Ozgurel, Pauzat,
  Ellinger, Maggiolo, Wurz, Vernazza, Lunine, {et~al.}}]{mousis2016origin}
Mousis, O., Ronnet, T., Brugger, B., {et~al.} 2016, \apj Letters, 823, L41

\bibitem[{Mumma {et~al.}(1972)Mumma, Stone, Borst, \&
  Zipf}]{mumma1972dissociative}
Mumma, M., Stone, E., Borst, W., \& Zipf, E. 1972, \jcp, 57, 68

\bibitem[{Nemeth {et~al.}(2016)Nemeth, Burch, Goetz, Goldstein, Henri,
  Koenders, Madanian, Mandt, Mokashi, Richter, {et~al.}}]{nemeth2016charged}
Nemeth, Z., Burch, J., Goetz, C., {et~al.} 2016, Monthly Notices of the Royal
  Astronomical Society, 462, S415

\bibitem[{Noonan {et~al.}(2016)Noonan, Schindhelm, Parker, Steffl, Davis,
  Stern, Levin, Kempf, \& Horanyi}]{noonan2016investigation}
Noonan, J., Schindhelm, E., Parker, J.~W., {et~al.} 2016, \actaa

\bibitem[{Sheel {et~al.}(2012)Sheel, Haider, Withers, Kozarev, Jun, Kang,
  Gronoff, \& Simon~Wedlund}]{sheel2012numerical}
Sheel, V., Haider, S., Withers, P., {et~al.} 2012, \jgr: Space Physics, 117

\bibitem[{Shirai {et~al.}(2001)Shirai, Tabata, \& Tawara}]{shirai2001analytic}
Shirai, T., Tabata, T., \& Tawara, H. 2001, Atomic Data and Nuclear Data
  Tables, 79, 143

\bibitem[{Stern {et~al.}(2007)Stern, Slater, Scherrer, Stone, Versteeg,
  A?hearn, Bertaux, Feldman, Festou, Parker, {et~al.}}]{stern2007alice}
Stern, S.~A., Slater, D., Scherrer, J., {et~al.} 2007, \ssr, 128, 507

\bibitem[{Stern {et~al.}(2015)Stern, Feaga, Schindhelm, Steffl, Parker,
  Feldman, Weaver, A’hearn, Cook, \& Bertaux}]{stern2015first}
Stern, S.~A., Feaga, L., Schindhelm, E., {et~al.} 2015, Icarus, 256, 117

\bibitem[{Taquet {et~al.}(2016)Taquet, Furuya, Walsh, \& van
  Dishoeck}]{taquet2016primordial}
Taquet, V., Furuya, K., Walsh, C., \& van Dishoeck, E.~F. 2016, \mnras, 462,
  S99

\bibitem[{Timar {et~al.}(2017)Timar, Nemeth, Szego, Dosa, Opitz, Madanian,
  Goetz, \& Richter}]{timar2017modelling}
Timar, A., Nemeth, Z., Szego, K., {et~al.} 2017, Monthly Notices of the Royal
  Astronomical Society, 469, S723

\bibitem[{Vatti~Palle {et~al.}(2004)Vatti~Palle, Ajello, \&
  Bhardwaj}]{JGRA:JGRA16941}
Vatti~Palle, P., Ajello, J., \& Bhardwaj, A. 2004, \jgr: Space Physics, 109,
  n/a, a02310.
\newblock \url{http://dx.doi.org/10.1029/2003JA009828}

\bibitem[{Vincent {et~al.}(2016)Vincent, A'Hearn, Lin, El-Maarry, Pajola,
  Sierks, Barbieri, Lamy, Rodrigo, Koschny, {et~al.}}]{vincent2016summer}
Vincent, J.-B., A'Hearn, M.~F., Lin, Z.-Y., {et~al.} 2016, Monthly Notices of
  the Royal Astronomical Society, 462, S184

\bibitem[{Vourlidas {et~al.}(2007)Vourlidas, Davis, Eyles, Crothers, Harrison,
  Howard, Moses, \& Socker}]{vourlidas2007first}
Vourlidas, A., Davis, C.~J., Eyles, C.~J., {et~al.} 2007, \apj Letters, 668,
  L79

\bibitem[{Witasse {et~al.}(2017)Witasse, S{\'a}nchez-Cano, Mays, Kajdi{\v{c}},
  Opgenoorth, Elliott, Richardson, Zouganelis, Zender, Wimmer-Schweingruber,
  {et~al.}}]{witasse2017interplanetary}
Witasse, O., S{\'a}nchez-Cano, B., Mays, M., {et~al.} 2017, Journal of
  Geophysical Research: Space Physics, 122, 7865

\end{thebibliography}
\end{document}